\documentclass[10pt,journal,compsoc]{IEEEtran}
\usepackage{comment}
\usepackage{ragged2e}
\usepackage{fontawesome5}
\usepackage{xcolor}
\usepackage{ulem}
\usepackage{enumitem}
\usepackage{multirow}
\usepackage{hyperref}
\setlist[itemize]{align=parleft,left=0pt..1.5em}

\usepackage{graphicx}
\usepackage{graphicx,subfigure}
\usepackage{wrapfig}
  \usepackage{cite}
\ifCLASSINFOpdf
\else
\fi
    \newcommand{\rqone}{\textbf{\textit{What factors are considered by the developers while self-assigning tasks?}}}
\newcommand{\rqtwo}{\textbf{\textit{How do managers facilitate sustainable self-assignment practices?}}}

\begin{document}
%
\title{What Drives and Sustains Self-Assignment \\in Agile Teams}
%
%
%
%

\author{Zainab Masood,~
        Rashina Hoda,~
        and~Kelly Blincoe
\IEEEcompsocitemizethanks{\IEEEcompsocthanksitem Z. Masood is with the Department
of Electrical, Computer, and Software Engineering, The University of Auckland, New Zealand.\protect\\
E-mail: zmas690@aucklanduni.ac.nz
\IEEEcompsocthanksitem R. Hoda is with Faculty of Information Technology, Monash University, Australia. E-mail: rashina.hoda@monash.edu 
\IEEEcompsocthanksitem K. Blincoe is with the Department of Electrical, Computer, and Software Engineering, The University of Auckland, New Zealand.}
\thanks{Manuscript received November 27, 2020}}
%
%

\markboth{Accepted for publication in IEEE TRANSACTIONS ON SOFTWARE ENGINEERING}%
{Shell \MakeLowercase{\textit{et al.}}: Bare Demo of IEEEtran.cls for Computer Society Journals}
%



\IEEEtitleabstractindextext{%
\justify
\begin{abstract}
Self-assignment, where software developers choose their own tasks, is a common practice in agile teams. However, it is not known why developers select certain tasks. It is important for managers to be aware of these reasons to ensure sustainable self-assignment practices. We investigated developers' preferences while they are choosing tasks for themselves. We collected data from 42 participants working in 25 different software companies. We applied Grounded Theory procedures to study and analyse factors for self-assigning tasks, which we grouped into three categories: \textit{task-based}, \textit{developer-based}, and \textit{opinion-based}. We found that developers have individual preferences and not all factors are important to every developer. Managers share some common and varying perspectives around the identified factors. Most managers want developers to give higher priority to certain factors. Developers often need to balance between task priority and their own individual preferences, and managers facilitate this through a variety of strategies. More \textit{risk-averse} managers encourage expertise-based self-assignment to ensure tasks are completed quickly. Managers who are \textit{risk-balancing} encourage developers to choose tasks that provide learning opportunities only when there is little risk of delays or reduced quality. Finally, \textit{growth-seeking} managers regularly encourage team members to pick tasks outside their comfort zone to encourage growth opportunities. Our findings will help managers to understand what developers consider when self-assigning tasks and help them empower their teams to practice self-assignment in a sustainable manner.
\end{abstract}

}

\maketitle

\IEEEdisplaynontitleabstractindextext

%
\IEEEpeerreviewmaketitle

\IEEEraisesectionheading{\section{Introduction}\label{sec:introduction}}

%
%
%
%
\IEEEPARstart{A}{gile} principles, values, and methods empower and enable autonomy in development teams. One of the principles behind the agile manifesto promotes \textit{trusting the individuals to get the job done}. Similarly, agile methods such as Scrum and Extreme Programming embed the value of \textit{respect}, which refers to the development team's right to receive authority and responsibility over their work. Agile teams tend to apply this empowerment and autonomy through \textit{self-organization}, \textit{self-management}, and \textit{self-assignment}, theoretically essential features of agile teams~\cite{hoda2016multi}. Agile team members have the freedom to self-assign tasks to themselves. However, there is limited literature on what factors developers consider and managers prefer developers to consider while self-assigning tasks. It is unknown what developers take into account while making self-assignment decisions when they have the power to choose their own tasks. Understanding this can give insights on what makes tasks attractive to practitioners and how sustainable self-assignment is practiced in agile teams.

The shift from traditional to agile software development has significantly changed the way task allocation takes place in teams. In traditional task allocation, the manager was responsible for assigning tasks, and studies have highlighted the factors managers consider when making task assignments. For example, we know that project managers and program managers of global software development projects consider developer expertise, task size, diversity of work, and autonomy when allocating tasks~\cite{imtiaz2017dynamics}. It is not known if developers also consider similar factors when they self-assign tasks to themselves.

This research aims to explore developers' individual preferences and factors they consider while self-assigning tasks. The study also explores the factors managers think their team members consider and those that the managers prefer their team members to consider while self-assigning tasks. The study investigates how managers balance/reconcile individual preferences and business needs to both keep the developers motivated and meet project goals. This research is guided by two main research questions:
\begin{itemize}
    \item \textbf{RQ1:} \rqone 
    \item \textbf{RQ2:} \rqtwo
\end{itemize}
\begin{table*}[]
\centering
\caption{Demographics of Participants}
\label{tab:participants}
\begin{tabular}{lllllll|lllllll} 
\hline
\textbf{P\#} & \textbf{Age} & \textbf{G} & \textbf{Role}        & \textbf{Domain} & \begin{tabular}[c]{@{}l@{}}\textbf{Total }\\\textbf{ Exp}\end{tabular} & \begin{tabular}[c]{@{}l@{}}\textbf{Agile }\\\textbf{ Exp}\end{tabular} & \textbf{P\#} & \textbf{Age} & \textbf{\textbf{G}} & \textbf{Role}    & \textbf{Domain} & \begin{tabular}[c]{@{}l@{}}\textbf{Total}\\\textbf{ Exp}\end{tabular} & \begin{tabular}[c]{@{}l@{}}\textbf{Agile}\\\textbf{ Exp}\end{tabular}  \\ 
\hline
P1           & 31-35        & M          & Developer            & IT              & 10                                                                     & 3                                                                      & P22          & 31-35        & M                   & Developer        & IT              & 11                                                                    & 4                                                                      \\
P2           & 31-35        & M          & Lead Developer & MD              & 13                                                                     & 7                                                                      & P23          & 46-50        & W                   & Tester           & FN              & 10                                                                    & 2.5                                                                    \\
P3           & 36-40        & M          & Team Lead; SM        & TP              & 17                                                                     & 7                                                                      & P24          & 36-40        & M                   & Head of Delivery & HC              & 13                                                                    & 3                                                                      \\
P4           & 31-35        & M          & Developer            & IT              & 10                                                                     & 6                                                                      & P25          & 36-40        & M                   & Developer        & RT              & 10                                                                    & 5                                                                      \\
P5           & 21-25        & W          & Developer            & ACC             & 2                                                                      & 2                                                                      & P26          & 31-35        & M                   & Dev Manager               & IT; CR          & 14                                                                    & 9                                                                      \\
P6           & 36-40        & M          & Architect            & ICT             & 10                                                                     & 3                                                                      & P27          & 36-40        & M                   & Scrum Master     & INV             & 12                                                                    & 4                                                                      \\
P7           & 21-25        & M          & Developer            & HC              & 2.5                                                                    & 1.5                                                                    & P28          & 50-55        & W                   & Tester           & RT              & 16                                                                    & 14                                                                     \\
P8           & 41-45        & M          & Team Lead; Developer       & IT              & 20                                                                     & 3                                                                      & P29          & 36-40        & W                   & Tester           & IT; PR          & 5                                                                     & 3                                                                      \\
P9           & 36-40        & W          & Scrum Master         & CR              & 9                                                                      & 6                                                                      & P30          & 40-45        & M                   & Manager          & IT              & 15                                                                    & 10                                                                     \\
P10          & 41-45        & M          & Developer            & HC              & 12.5                                                                   & 6                                                                      & P31          & 26-30        & M                   & Scrum Master     & IT              & 4                                                                     & 1.5                                                                    \\
P11          & 31-35        & M          & Tester               & FN; BK          & 10                                                                     & 5                                                                      & P32          & 41-45        & M                   & Scrum Master     & IT              & 20                                                                    & 12                                                                     \\
P12          & 31-35        & W          & Tester               & MD              & 12                                                                     & 1                                                                      & P33          & 31-35        & M                   & Product Owner    & IT              & 13                                                                    & 8                                                                      \\
P13          & 31-35        & M          & Developer; Tester    & HC; BK          & 10.5                                                                   & 4                                                                      & P34          & 26-30        & M                   & Developer        & NT              & 4                                                                     & 3                                                                      \\
P14          & 31-35        & M          & Product Owner        & IT; TC          & 12                                                                     & 5                                                                      & P35          & 36-40        & M                   & Scrum Master     & IT              & 11                                                                    & 3                                                                      \\
P15          & 36-40        & M          & Developer            & HC              & 12                                                                     & 2                                                                      & P36          & 26-30        & W                   & Business Analyst & IT;RM           & 9                                                                     & 6                                                                      \\
P16          & 26-30        & M          & Developer            & IT              & 4                                                                      & 3.5                                                                    & P37          & 31-35        & M                   & Architect        & IT;RM           & 13                                                                    & 4                                                                      \\
P17          & 31-35        & M          & Team lead; SM        & IT              & 8                                                                      & 3.5                                                                    & P38          & 21-25        & M                   & Tester           & NT              & 2                                                                     & 2                                                                      \\
P18          & 46-50        & M          & Team Lead; Developer       & IT              & 25                                                                     & 9                                                                      & P39          & 26-30        & W                   & Scrum Master     & IT              & 3                                                                     & 1                                                                      \\
P19          & 46-50        & M          & Dev Manager                   & HR              & 20                                                                     & 2                                                                      & P40          & 31-35        & M                   & Dev Manager               & RT              & 14                                                                    & 9                                                                      \\
P20          & 36-40        & M          & Developer            & IT              & 12                                                                     & 7                                                                      & P41          & 31-35        & M                   & Technical lead   & IT              & 10                                                                    & 5                                                                      \\
P21          & 31-35        & M          & Tester               & MD; HC          & 10                                                                     & 3                                                                      & P42          & 31-35        & M                   & Manager          & IT              & 10                                                                    & 5                                                                      \\ 
\hline
\multicolumn{14}{l}{Participant P\#, Gender G, Experience Exp, Role [SM=Scrum Master; Dev Man=Development Manager];Domain [IT=Information Technology;}\\
\multicolumn{14}{l}{ ICT=Information  Communication Technologies; MD=Medical; TP=Transport; ACC=Accounting; FN=Finance, HC=Healthcare; RT= Retail; }\\
\multicolumn{12}{l}{ NT=Networking; PR=Payroll; RM=Requirements Management; BK=Banking]} \\ 
\hline
\end{tabular}
\end{table*}

We conducted a large Grounded Theory (GT) study with 54 participants from 26 software companies on the phenomenon of \textit{self-assignment in agile software development teams} using the Strauss-Corbinian version of GT~\cite{straus1990basics}. We applied the full GT method, including interleaved rounds of data collection and analysis procedures, such as open coding, axial coding, and constant comparison. We identified a number of key findings, including two that help answer the above research questions and are reported in this paper, \textit{self-assignment factors} and \textit{manager strategies}. The self-assignment factors present the final in-depth findings based on the full GT study, and build on the preliminary list of motivating factors identified through a pilot study, reported in a short paper~\cite{masood2017motivation}. The other main finding of the wider GT study was a theory of \textit{how self-assignment works in agile teams} \cite{masood2020agile}.

The contributions of the paper are as follows. Firstly, we describe a set of factors developers consider during self-assignment. Secondly, we compare the factors from the developers and managers perspectives and report the variations between them. Thirdly, we report the strategies managers use to balance/reconcile individual preferences and business needs. Most importantly, the paper provides a set of guidelines as recommendations for managers to ensure that all developers are provided opportunities to learn new technology, tools, and techniques in practice. 
\begin{figure*}[t]
\includegraphics[scale=0.545]{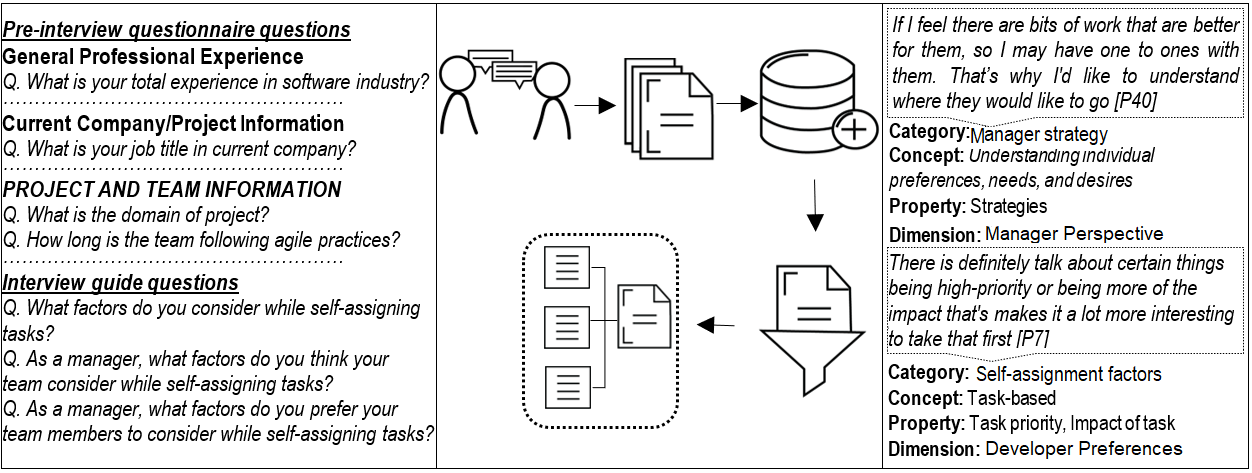}
\centering \caption{Data Collection \& Analysis Process (Left box: Sample questions from pre-interview questionnaire \& interview guide, Middle box: Interviews, Transcription, Data Management [NVivo], Data Extraction, Data Analysis, Right box: Strauss-Corbinian GT analysis procedure: Open Coding)}
  \label{fig:process}
\end{figure*}
\section{Background}
\label{sec:relwork}
In traditional software development, managers were responsible for selecting and allocating tasks. Decisions around allocating tasks to the right people within a team are not easy. These decisions, if done right, can increase the chance of successful project completion~\cite{demarco1999peopleware}. Studies have explored factors that managers take into account while making task allocation decisions. A study of global software development teams found that project managers consider \textit{task size}, \textit{autonomy}, \textit{expertise}, and \textit{variety of work} when allocating tasks to developers~\cite{imtiaz2017dynamics}. Another study found that managers consider \textit{business expertise} and the \textit{developer’s technical knowledge} when allocating tasks in global software development~\cite{simao2016task}. 

On the contrary, in agile software development, teams have the freedom to choose their own tasks, theoretically enabling \textit{self-assignment}, a key practice of agile software development~\cite{deemer2012lightweight, hoda2016multi}. However, in practice, task allocation in agile teams happens through different strategies: manager-driven, team-driven, manager-assisted, team-assisted, or individual-driven~\cite{masood2017exploring}. Team members have different individual preferences when they self-assign tasks. Nevertheless, the number of studies on this topic is scarce, existing studies investigated developer work motivation in general~\cite{hall2008we, beecham2008motivation, francca2011motivation} and did not specifically focus on factors in task allocation in agile software development contexts.

As part of the pilot study with 12 developers, we identified an initial list of factors that are considered while self-assigning tasks, under three groups: task-based, developer-based, and opinion-based~\cite{masood2017motivation}. Out of these, task-based and developer-based factors were found to be considered more important by the developers. The results of this paper present the final and mature findings that build on the pilot study. In this paper, we describe the mature findings from the full GT study with 42 practitioners from agile teams in 25 companies (not including the 12 participants from the pilot study). This in-depth study confirms the initial factors from our pilot study and identifies seven additional factors. It also extends the knowledge with the addition of managers' preferences and perspectives around developers' preferences while self-assigning tasks, presented as \textit{manager strategies and approaches}, filling the literature gap on factors behind self-assignment choices in agile teams and presenting `both sides of the story'.

\section{Research Methods}
\label{sec:methods}

We conducted a large GT study, following the Strauss-Corbinian guidelines~\cite{straus1990basics}. Details of the interleaved data collection and analysis are presented in this section. As with any large GT study, it is common to identify a number of key findings and report them in different, detailed manuscripts. One of the key findings of our wider study, a theory of \textit{how self-assignment works}, has been published elsewhere~\cite{masood2020agile}. This paper presents other key findings related to what drives and sustains self-assignment in agile teams, i.e., \textit{self-assignment factors} and \textit{manager strategies}.

\subsection {Semi-structured Interviews}
We collected data through semi-structured, face-to-face interviews. The authors collectively drafted an interview guide focusing on the research outcomes. Some of the sample questions from the interview guide are listed in Figure \ref{fig:process}. The interview questions in the guide were kept open-ended to encourage discussions. This guide was modified over the research period with cycles of data collection, analysis, and theoretical sampling. For example, in the initial round of interviews we asked the developers \textit{`What factors do you consider while self-assigning tasks?'} and managers \textit{`What factors do you think your team members consider while self-assigning tasks?'} for triangulation. Based on the initial analysis of these responses, we added additional questions for the following rounds of interviews: \textit{`What factors do your managers prefer you to consider while self-assigning tasks?'}, \textit{`What factors do you prefer your team members consider while self-assigning tasks?'} and \textit{`How do you facilitate self-assignment to reconcile individual preferences with business priority and project needs?'}. All of the interviews were audio recorded, transcribed verbatim, and saved in NVivo, a popular data analysis software tool.

\subsection{Recruiting Participants}
We sent invitations to multiple agile practitioners using social networking sites such as LinkedIn and Meetup groups. Those who got back to us showing willingness to participate were requested to share their details using a pre-interview questionnaire, which captured the demographics of the participants such as basic information, professional details, and agile experience using Google Forms. This helped to get insights into the agile practices followed by each potential participant and enabled us to select only participants who practiced self-assignment at least occasionally. 

\subsection{Participants Demographics}
We interviewed 42 agile practitioners who were working for software companies across various domains such as IT, healthcare, transport, accounting, and finance. The primary roles of the participants were Software Developers (n=11), Testers (n=7), Architects (n=2), Scrum Masters (n=7), Managers (n=6), Business Analyst (n=1), and Product Owners (n=2). Several participants held more than one role, e.g., Developer and Tester (n=1), Team lead and Scrum Master (n=2). All of the participants were working on projects employing agile methodologies and using key agile practices such as iteration planning, daily stand-up, reviews, and retrospectives. A majority of the participants were practicing Scrum, but a few used a combination of Scrum and Kanban. The participants varied in gender, age, and professional experience. Their software experience ranged from 2 to 25 years, and their ages ranged from 21 to 55 years. Participants of the study are represented by numbers from P1 to P42 to maintain the participants' anonymity (see Table \ref{tab:participants}). In the remainder of the paper, we use the term \textit{'manager'} to refer to participants with management roles (project manager, scrum masters, product owners, team leads). 
 We use the term \textit{'dev'} to represent all other roles in the development team such as testers, developers, test analysts, architects, and business analysts. 

\subsection{Data Analysis }
We  applied Strauss and Corbin's GT data analysis procedures, i.e., open coding, constant comparison, and axial coding on our dataset~\cite{straus1990basics}. Figure~\ref{fig:process} illustrates the application of analysis procedures on two raw interview transcripts of the participants [P7, P40].

During open-coding, we analysed the data and labeled it with short phrases (two to three words) as codes. Then, we grouped similar codes to a higher level of abstraction called concepts. The concept of 'task-based' factor emerged from the transcript of participant P7 (See Figure~\ref{fig:process}). Similarly, the concept of 'understanding individual preferences and needs' emerged from the transcript of P40. These concepts were defined in terms of their properties and dimensions to refine them further.\\

Properties are ‘characteristics that define and explain a concept’. For example, two different properties of the task are identified from the transcript of participant P7 as 'task \textit{priority}' and 'task \textit{impact}'. Similarly, from the transcript of participant P40, ‘strategy’ emerged as a property. Dimensions are ‘variations within properties’. For example, we considered variations in managers' and developers' perspectives and preferences in this study. Figure~\ref{fig:process} shows examples of a manager's [P40] perspective and a developer's~[P7] preferences. Another example of a property is '\textit{previous experience}', which further explains the concept of developer-based factor and considers various dimensions, such as previous experience with tools, skills, technologies, domain, and nature of work (e.g., dependent tasks or related stories/tasks). Other similar concepts and properties were identified and grouped under the developer-based factors through constant comparison during open-coding. Similarly, `opinion of team members' included multiple concepts such as not meeting team members' expectations, suggestions of peers, and suggestions of senior team members. We grouped all of these concepts under opinion of team members.

Open-coding was applied on the entire data set, resulting in a set of concepts and categories. For example, we combined related properties under the concepts \textit{task-based} factors, \textit{developer-based} factors, and \textit{opinion-based} factors, which combined to a category of self-assignment factors. Figure~\ref{fig:process} demonstrates how open coding led to the emergence of the categories 'self-assignment factors' and 'manager strategy'.

These categories, concepts, properties, and dimensions were developed and improved over time with ongoing discussions and constant feedback from the co-authors. Some of these concepts and categories were similar to our preliminary study (e.g., opinion-based factors)\cite{masood2017motivation}. Some new properties were added to the previous concepts (e.g., task-based factors, developer-based factors) maturing our category of self-assignment factors. New concepts and categories also evolved during the analysis procedure (e.g., manager strategies, manager approaches).

We applied axial coding to identify relationships in our data. For example, axial coding uncovered that a developer’s tenure could impact their self-assignment preferences and it uncovered the trade-offs that team members make in order to work on tasks with learning potential. These and other relationships identified during the axial coding are described in Section~4.3. We identified relationships not just within the sub-categories but also across the different sub-categories. We validated these relationships on the entire data set. This involved reviewing data iteratively and discussing the potential relationships within the research team. We analysed the data until theoretical saturation was reached, i.e., no new concepts, categories, or relationships were identified with further data collection on self-assignment factors. This paper presents a detailed account of key categories that emerged, namely, 'self-assignment factors', 'managers’ balancing strategies', and 'managers approaches'.

\section{\rqone (RQ1)}
\label{sec:rq1}
We found that developers consider various factors while self-assigning tasks (presented in Table~\ref{tab:factors}).

\subsection{Self-assignment Factors}
We grouped the factors into three sub-categories: task-based, developer-based, and opinion-based factors. Here we describe each factor and provide representative quotes. 

\subsubsection{Task-based Factors}
The task-based factors are primarily related to different attributes of the task itself, such as the business priority, technical complexity, or visibility.\\

\noindent \textbf{\textit{Task business priority}} refers to how important and urgent a task is to the business and customer. A task with higher priority is more important than a task with lower priority. The highest priority tasks are meant to be at the top of the backlog and completed before tasks of lower priority. Thus, team members typically choose the highest priority tasks.\\
\textit {Priority is very important thing because we line up stories with priorities on board and we need to pick ones at the top first. \textbf{P5~(dev)}} \\\\ 
\textbf{\textit {Task learning potential}} refers to the potential for the task to provide an opportunity to learn new skills or explore a new area of interest. Some tasks have more room for learning new technologies, tools, or domains.\\
\textit{There's a lot more excitement about working on tasks with kind of newer technologies, new frameworks and stuff. \textbf{P17~(manager)}}\\\\
\textbf{\textit{Technical complexity}} has to do with how difficult a task is. Some tasks can be more difficult than others based on the technology involved, the amount of code that needs to be changed, or the complexity of the logic or algorithms.\\
\textit {For me [a factor is] how challenging a task is. \textbf{P20~(dev)}}\\
\textit {I avoid assigning myself to a complex task and then not be able to deliver some outcome. \textbf{P5~(dev)}}\\
Some developers prefer complex tasks, while others prefer easy tasks.  This preference also varies based on other factors. For example, if a developer already has a high workload (\textit{Resource Availability} factor), they may prefer to take only easy tasks.\\\\
\textbf{\textit{Task visibility and impact}} refers to a task's potential for recognition compared to other tasks. This means there are some tasks that, when accomplished, are more appreciated by the team leader or department head. Team members prefer to choose tasks which have more visibility and impact.\\
\textit {Like if one's [task] a data entry or one's data fix and one's fixing like a bill pipeline or doing something like that, then obviously, the bill pipeline one will win because it's high benefit to them [customer], or high visibility, and doing it feels good. \textbf{P2~(dev)}}\\
However, the most complex or large tasks are not always the most impactful. A tiny change can be very impactful as stated by one participant.\\
\textit {If you can do this one little feature which when the head of our department sees it, he's going WOW! \textbf{P2~(dev)}}\\\\
\textbf{\textit{Task dependencies}} refers to relationships with other tasks, teams, or external entities. Some tasks will rely on other tasks to be completed before they can be started, and some will require coordinating with other teams working on related tasks. Some tasks could even require dealing with third parties such as other companies. \\
\textit{We've often got some dependent cards and it makes sense that they are picked by someone who worked on the related card. \textbf{P2~(dev)}}\\\\
\textbf{\textit{Completion time}} is how quickly the individual can accomplish that task. Team members often pick up tasks they think they can finish quickly.\\
\textit{Can I do it [task] quickly? \textbf{P10~(dev)}}\\
Sometimes, individuals will let someone else take a task if they believe the other person can get things done faster.\\
\textit{I've done this 20 times before, it's easy for me, taking only five minutes, I'll do it.... the other developer in my squad, if he's very experienced in it I might say, you know, he can do this much quicker, then I'll let him do it. \textbf{P1~(dev)}}\\
It can also be inferred from the above quote that completion time is related to prior experience, e.g., having prior experience in the problem space or technology allows an individual to complete tasks more quickly than someone with less experience.\\\\
\textbf{\textit{Understandability}} of the problem domain refers to how well the individual understands the problem to be addressed in the task. \\
\textit{If someone [team member] doesn't understand, they are less inclined to put their hands up coz you need to feel confident that you can implement it. \textbf{P9~(manager)}}\\
Tasks can be difficult to understand because of missing details, as reported by one participant.\\
\textit{So there’s often tasks, they’re not very well [explained] hard to understand. \textbf{P2~(dev)}}\\ 
Another participant noted that implementing features for other teams can be challenging in terms of understanding since they may not provide all needed information.\\
\textit{Supporting other teams that don't have enough information... they won't tell us what it does and say you have to do this and this and that doesn't mean much. \textbf{P7~(dev)}}\\\\
\textbf{\textit{Task desirability}}  has to do with the pleasure of a task. Tasks that are enjoyable will be selected before boring or dull tasks. For example, as noted by P2: \\
\textit{There's definitely an aspect of fun, if card [task] is more fun than another card then it will obviously get a higher priority because there's more personal enjoyment associated with it. \textbf{P2~(dev)}}

\subsubsection{Developer-based Factors}
Developer-based factors are the factors associated with the developer who is selecting the task (e.g., their previous experience, technical expertise, or co-worker preferences) or their desire to provide opportunities for other team members. \\\\
\textbf{\textit{Previous experience}} has to do with the past tasks an individual has worked on. Having worked on similar tasks in the past may make a task more appealing. This factor considers past experience with the tools and technologies involved with the task and also the nature and domain of the task (i.e., front-end, back-end tasks, Android or iOS related tasks, programming languages, etc.).\\
\textit{If it's like an investigation that turned into story and I did that [investigation] part. \textbf{P7~(dev)}}\\
\textit{I'd say developers that are more senior in team, have been around longer and have familiarity with some of our legacy systems, they actually gravitate towards those, more the legacy tickets. \textbf{P17~(manager)}}\\
This factor contradicts the \textit{task learning potential} task-based factor. Self-assigning tasks with prior working experience reduces the opportunity to learn new technology, tools, or domains. 
\\\\
\textbf{\textit{Technical expertise and ability}}  is the competency of the individual to perform tasks based on their knowledge, abilities, and technical skills.\\
\textit{If I have got technical expertise in that area. \textbf{P7 ~(dev)}}\\
\textit{People do have their own sort of strengths and tend to pick up the work that relates to their strengths. \textbf{P16~(dev)}} \\
\textit{whether I am capable I mean technically able to do that \textbf{P5~(dev)}}\\
This factor is different from \textit{previous experience}. For example, a php expert may not have any previous experience of building a shopping cart, but they have the required technical skills. \\\\
\textbf{\textit{Resource availability}} refers to the schedule and workload of a team member. If someone’s schedule is full or their workload is high, they may not be willing to pick up complicated tasks, which will require more time to understand and develop. \\
\textit{I think possibly how pressurized it [task] is in our current workload. \textbf{P9~(manager)}}\\
Similarly, agile practitioners also consider the context switches that are required based on their current workload. For example, one participant noted that:\\
\textit{You're always coming off something and moving on to something else. The thing that you are coming off if the end isn't clear in sight, you are less inclined to want to take on another complex [task] because you need to context switch.... I want to get through this complex issue before I move on to the next complex issue. \textbf{P9~(manager)}}\\\\
\textbf{\textit{Co-workers deference}} is seen for peers and juniors. \textit{Deference to peers} refers to situations where individuals voluntarily avoid a task to give their peers an opportunity to choose it first.\\
\textit{There have been lots of times where I thought of that but you know it's a bit unfair to sort of take everything that you want and normally other people want. \textbf{P7~(dev)}}\\
\textit{Deference to juniors} refers to a similar situation when the senior team members intentionally let junior team members select tasks first.\\
\textit{There have been cases where I [lead] kind of hold back. I don't pick any of the stories, and I let the other guys [juniors] pick what they want first. And then I'll do just anything which is left over.~\textbf{ P8~(manager)}}\\\\
\textbf{\textit{Co-workers preference}} refers to the aspiration of a team member to work with certain people.\\
\textit{The people you are going to work with, that influence while you self-assign, because I like to work with developer [X], so I will probably pick up a ticket that X will be working on. \textbf{P14~(manager)}}
\subsubsection{Opinion-based Factors} We found that the opinions of managers and team members are considered in some situations when self-assigning tasks.
\textbf{\textit {Opinion of Managers}} means that the developer considers their manager's opinion when selecting tasks.\\ 
\textit{My [Manager's] presence influences their calls, ... and felt that I could be a little bit coercive too by saying yeah, X would be best to work on that one, and then suddenly he's assigned to it because I said that. \textbf{P19~(manager)}} \\\\
\textbf{\textit{Opinion of other team members}} is when the developers consider the opinions of their co-workers when selecting a task. They want to be regarded as being respectful and helpful by their team members.\\
\textit{You don't want to disappoint your peers, you don't want to be doing something which isn't high value. You don't want to be caught out on doing something rubbish. \textbf{P16~(dev)}}

\subsection {Developers' Preferences Vs Managers' Perspectives Vs Managers' Preferences}

Here, we examine the factors that developers prefer (\textit{developers' preferences}), the factors managers think developers consider  (\textit{managers' perspectives}), and the factors that managers
prefer (\textit{managers' preferences}). These preferences are shown in Table~2. 

\subsubsection{Developers' Preferences}
We  examined  the  commonality  of  the  reported  factors based on the number of developers who mentioned them in our dataset. Our results indicate that the factor \textit{task learning potential} is the most commonly reported factor by the developers. The second most common factor reported by the developers is \textit{technical complexity}. Other common task-based factors are \textit{business priority}, \textit{task dependencies}, and \textit{understandability}.

From the developer-based factors \textit{previous experience} and \textit{technical expertise and ability} are the two most commonly reported factors, showing that developers prefer to choose tasks in line with their past experience and their skill set. Their experience and technical expertise makes these tasks easier to accomplish, which helps the development process run smoothly and improves productivity. Other common developer-based factors are resource availability, co-workers deference, and co-workers preference.

For the opinion-based factors, considering the opinion of managers is more common than considering the opinion of other team members. Compared to task-based and developer-based factors, opinion-based factors are considered less by the developers.
\begin{table} \label{tab:factors}
\centering
\caption{Developers' Preferences Vs Manager' Preferences of Self-assignment Factors}
\begin{tabular}{llcc} 
\hline
\textbf{Category~}      & \textbf{Self-Assignment Factors} & \begin{tabular}[c]{@{}c@{}}\textbf{\begin{tabular}[c]{@{}c@{}}Devs\\ Prefs\end{tabular}}\end{tabular} & \textbf{\begin{tabular}[c]{@{}c@{}}Mgrs\\ Prefs\end{tabular}}\\
\hline
\multirow[t]{8}{*}{Task-based}      & Task's business priority      & \faCheck    & \faCheck  \\
                               & Task learning potential       & \faCheck    & \faCheck \\
                               & Technical complexity          & \faCheck    &      \\
                               & Task visibility  impact       & \faCheck    &      \\
                               & Task dependency               & \faCheck    & \faCheck \\
                               & Completion time               & \faCheck    & \faCheck  \\
                               & Understandability             & \faCheck    & \faCheck  \\
                               & Task desirability             & \faCheck    &      \\
\multirow[t]{5}{*}{Developer-based} & Previous experience           & \faCheck    & \faCheck  \\
                               & Technical expertise  ability  & \faCheck    & \faCheck  \\
                               & Resource availability         & \faCheck    &      \\
                               & Co-workers deference          & \faCheck    &      \\
                               & Co-workers preference         & \faCheck    &      \\
\multirow[t]{2}{*}{Opinion-based}   & Opinion of managers           & \faCheck    &      \\
                               & Opinion of team members       & \faCheck    &    \\ \hline
\multicolumn{4}{l}{Preferences \textbf{Prefs}, Developers' \textbf{Devs}, Managers' \textbf{Mgrs}} \\
\hline
\end{tabular}
\end{table}
\subsubsection{Managers' Perspective}

We asked managers what they think developers consider while self-assigning tasks. Our results indicate that managers believe that developers consider the \textit{task learning potential} the most. Managers also believe technical complexity and previous experience were commonly considered factors by the developers, which is in line with the preferences reported by the developers.

However, in some cases, managers' perspectives were not in line with the preferences reported by developers. While we found that many managers think developers consider business priority significantly, this was not commonly reported by the developers. One potential reason for fewer developers reporting \textit{business priority} could be that most of the agile teams consider \textit{business priority} as an implicit factor. Agile teams are meant to deliver whatever is important to the customer. 
Managers also believed that the majority of the developers consider \textit{task dependencies} and \textit{understandability} at a higher rate than was reported by the developers. Managers also viewed the opinion-based factors to be considered by developers at a higher rate than the reported developers' preferences (\textit{opinion of managers} and  \textit{opinion of other team members}). 

\subsubsection{Managers' Preferences}
We also asked the managers what they prefer the developers to consider when self-assigning tasks, and we found that managers give precedence to \textit{quick completion} and \textit{previous experience}. They are happy if developers self-assign tasks related to their previous work and tasks they can complete quickly. For example, if a bug is reported, managers would prefer a developer who worked on the feature initially since they would have the best knowledge to resolve the bug. If the task is a high priority, customer-facing enhancement or any primary feature, the manager's preference would be someone with better domain knowledge, technical expertise, and prior experience as this increases the manager's confidence in quick and productive delivery. 

We noticed an interesting contrast between the factors that managers would like the developers to consider and those developers actually prefer. In practice, developers consider the task-based factors \textit{task learning potential} and \textit{technical complexity} the most. Some managers acknowledged and pointed out the variations between individual and manager preferences as: \\
\textit{So time is the most important factor[how quickly this needs to be done]. Previous experience of having worked in that area is another one. If those are not constraints, then we look for opportunities to learn. From my team's perspective, it’s the other way around. So opportunities to learn is something that is topmost from them. Second thing is everyone likes a challenge, so if there's something which is very complex, hard to deal with, or something, everyone wants to do it. And then anything that is under time pressure, people generally avoid it.} \textbf{{P24~(manager)}}

Another manager shared the difference of opinion as:\\
\textit{No, if I'm doing that [assigning] my motivation would be different. If somebody can do a better job more efficiently, more quickly, then definitely him because we [services company] bill for our time. \textbf{P33~(manager)}}

\textit{Previous experience and technical ability} are the common factors which developers are seen to consider and the managers want developers to keep into consideration while self-assigning tasks. Most of the  managers are  more confident and happy when developers self-assign tasks they understand and are good at doing.

\subsection{Developers' preference of factors and trade-offs} \label{sec:tradeoffscontext}
We examined how developer preferences vary under different contexts and situations. We found large variations based on the context. For example, a developer's tenure could impact their self-assignment preferences. Newly hired team members are more likely to pick tasks in their area of expertise that can be completed quickly so that they can establish a reputation on the team. They are less likely to choose tasks that will allow them to learn new technologies, tools, or domains for personal career growth. As another example, a developer mentioned that they self-assign tasks that require less assistance and support from others when they need to work from home.
\begin{table}[]
\caption{Developers' preference of factors and trade-offs}
\label{tab:tradeoffs}
\centering
\begin{tabular}{ll}
\cline{1-2}
\textbf{Developers prefer these factors } & \textbf{Over these factors}           \\ \cline{1-2}
Task learning potential  & Business priority   \\
Task dependency  &  Business priority           \\
Completion time    & Business priority          \\
Task desirability   & Business priority        \\
Task learning potential  & Understandability  \\
Task learning potential    & Completion time   \\
Task learning potential    & Previous experience   \\
Task learning potential    & Technical expertise \& ability   \\
Managers opinion    & Task desirability    \\ \cline{1-2}
\end{tabular}
\end{table}

Interestingly, we saw cases where some factors seem to be considered together. This was seen not only between the factors within a group but also across factors from different groups. For example, a task with low \textit{business value} and longer \textit{completion time} (both task-based factors) is less likely to be selected as stated by one of the participants below. 
\textit{Like if there's low business value [priority] in a task and it's going to take someone [team member] a long time there's not really much incentive to do something. \textbf{P2~(dev)}}

We found that developers ignore certain factors over others. Table~\ref{tab:tradeoffs} shows pairs of factors where the factor in column 1 is given preference over the associated factor listed in column 2. As can be seen, it is common for developers to give less importance to a task's \textit{ business priority}. For example, a participant stated that they had a team member who was fond of doing user interface tasks (\textit{task desirability} factor) and would select them regardless of their low \textit{business priority}. If someone has worked on a feature, they are more inclined to do all of the related work like future bug fixes or related enhancements (\textit{task dependency} factor). Similarly, one participant described ignoring a high \textit{business priority} task to pick up a task that was similar to the last one they completed quickly.\\
\textit{I've been working on iOS and I would prefer to pick up iOS cards [tasks]. Because it's all fresh in my mind, all the development rules. There has been a higher priority card available, but I have picked the iOS card because it [iOS task] makes sense for me to do it, coz I can finish it relatively quickly. \textbf{P11~(dev)}}

It can also be seen in Table~\ref{tab:tradeoffs} that a \textit{task’s learning potential} is often considered over other factors, ignoring the fact that the task may take longer (\textit{completion time}), may not be \textit{understandable}, or that the developer lacks \textit{previous experience} or the required \textit{technical expertise}.

Team members make trade-offs in order to work on tasks with learning potential, e.g., learning a new technology. They will put in extra effort, such as working extra hours, since it as an opportunity to grow and improve their skill set.
Individuals also pair-up to work on the task with another experienced team member (developer-based factors: \textit{preferred co-workers} and \textit{previous experience}). With this strategy, they are allowed to learn without delaying the task. \\
\textit{If I'm [developer] interested to learn I'll just sit with him [experienced developer], and see how he does it…. But I wouldn't pick something knowing that I have zero knowledge and it will take me two days and it'll take him 10 minutes to do.  I wouldn't say I'll do it because I'm interested, I'd say you do it but then I can sit with you and learn. \textbf{P1~(dev)}}

Even though developers prefer learning opportunities (task-based factor: \textit{task learning potential}) over tasks they can complete quickly (task-based factor: \textit{completion time}), they cannot always select tasks that would give them learning opportunities due to time constraints as shared by P14.\\
\textit{Sometimes organizations do not have time to deliver stuff, they want the main experts to deliver something. But you say, no, no, I [team member] want to learn something new, so I want to take it. \textbf{P14~(manager)}}\\
Other times, team members have to learn new technologies they are not interested in to ensure a task is delivered. Individuals are not always happy with letting business priority override their professional development.\\
\textit{Sometimes organizations want you [developers] to learn new stuff, but you still want to work e.g., in Java, because it's your language preference. \textbf{P14~(manager)}}
\\\\
\fbox{\begin{minipage}{26em}
\textbf{Summary for RQ1:} Developers consider different developer-, task-, and   opinion-based factors while self-assigning tasks. Managers are aware of some of these developer preferences, but at times overestimate the importance developers give to some of the factors. Not all factors are important to every developer,  and their  preferences often vary based on context. Developers make trade-offs to settle conflicting priorities and preferences, but they need  managers to facilitate this at times.
\end{minipage}}
\section{\rqtwo (RQ2)}
\label{sec:rq2}
In RQ1, we found tension between individual preferences, specifically the \textit{task learning potential} and \textit{business priorities}. Here, we investigate how managers balance providing opportunities to learn new technologies, tools, and domains without impacting the business priorities and project outcomes. We found that managers play a significant role in reconciling this tension.

\subsection{Managers' Balancing Strategies}
\label{sec:strategies}
We found managers use various strategies to balance developers' individual preferences with business needs. These balancing strategies are applied as per the context, and are summarised below with example quotes.

\textbf{\textit {Suggest tasks that align with individual preferences: }} 
The managers want developers to be aware of their individual preferences and personal goals and to self-assign tasks that align with both their personal goals and team goals. A personal goal could be learning a specific language, tool, or technology, and a team goal refers to what the team collectively aims to accomplish by the end of the sprint, release, or project. Managers encourage developers to share their personal goals with both the manager and the team so that they can help the developers achieve their personal goals while still contributing to the team goals. A manager shared an example where a team member shared their personal goal, and the manager helped point out tasks that aligned with that goal.\\
\textit{... I [manager] sort of push them to say 'hey perhaps you might be interested in doing this one, this is gonna be a quite interesting piece of work if you wanna learn x, y, and z'.  And so it's a combination of them picking their own based on their own professional development, as well as me saying perhaps you want to have a look at this one [matching to personal goals]. \textbf{P40~(manager)}}

\textbf{\textit {Rearrange tasks to align to individual preferences:}} 
As another strategy, managers often rearrange the iteration task list to provide learning opportunities to developers. A manager shared an example where developers worked on a product that was developed in both the Python and Go languages. The team had a few Go developers who would self-assign all of the Go tasks.  The manager noticed this and intervened to ensure that the GO tasks could be spread across the team. The manager rearranged the task list to include some easy Go tasks suitable for a novice Go developer and encouraged developers who wanted to learn Go to select them. This provided opportunities to learn and also acted as a gentle reminder to the Go experts to share their expertise to help others learn.\\
\textit{...... We had a lead who is also very good at Go.  As we were discussing who wanted what work they were picking some of the Python stuff, and he was getting the Go stuff. And I said, 'Well this is all good (laughter), but what we need to do is spread this Go work out a bit more'. We arranged the easier parts of the Go work. It was challenging them a little bit, but not throwing them in the deep end and so we did some sort of just rearranged the tasks to give them a nice introduction.} \textbf{P40~(manager)}

\textbf{\textit {Provide alternate learning opportunities: }} 
There are times when developers show interest in working on tasks where they have no prior experience or expertise. This becomes challenging for the managers. An interesting strategy to handle this is \textit{asking the developers to provide proof of skills}. One of the managers shared an example of a UI designer with no UX experience who wanted to self-assign to a UX task. The manager offered them an opportunity to learn and prove their interest by taking an introductory course, giving a demo to the team, or critiquing another developer's work and proposing a better alternative. The developers do not always accept these alternate learning opportunities. \\
\textit{.. We said 'Okay, how about you let go of this task in this sprint, don't dive into the UX. Complete a UX course first, maybe a one week, two weeks course, or give us a demo for what you can do with the user experience or try to critique X's work and see if you can come up with a better solution or if you can come up with better natural user flows,...with better, lesser number of tasks for one action, the good UX parameters, in the next sprint we can think about giving you the UX task.} \textbf{P33~(manager)}

This contrasts with the previous example where the developers did not explicitly state their goal of learning Go. It was the manager who noticed and intervened. However, here it was the developer who stated they wanted to learn UX. This highlights that teams are comprised of developers with different personalities, and managers need many different strategies.  

\textbf{\textit {Enable individual skills development after other tasks:}}
Another strategy is to allow developers to take time to learn after they have finished all of their other work in any remaining time left in the sprint. This provides an opportunity to the developers to learn new technologies, skills, and tools.  \\
\textit{If you finish your Sprint work early, do training, or learn something you want to} \textbf{P9~(manager)}

\textbf{\textit {Monitor individual skill development through mentoring: }}
Some managers assign mentors to every developer in the team. Together they define a learning path to ensure the developer's long-term growth, and it is the mentor's responsibility to monitor learning progress. The developers acknowledged that sometimes it is hard to manage learning activities with a routine workload, but that mentors can help them stay motivated. Interestingly, if a developer does not make any progress, then the mentor is held accountable, not the developer. The developers acknowledged this as a productive activity that helps them to improve their profile.

\textit {Discouraging failure (or embarrassment):}
Some managers were seen to discourage developers from self-assigning tasks where the risk of failing to complete it was high due to lack of required expertise. This was further enhanced by the awareness that certain tasks were considered high profile, with senior managers maintaining interest and visibility into their status. Managers were hesitant to let developers without required skills self-assign such high-profile tasks and actively discouraged them to do so to avoid embarrassment in front of senior management and any potential negative consequences arising from it.\\
\textit{But this task has high visibility, and if you fail at this task, management will notice. And you’re not going to do it perfectly, or I suspect you won’t do it perfect, and management is going to notice this.} \textbf{P26~(manager)} 

\medskip 

While managers have many strategies to accommodate developer preferences, they also want team members to understand that there will always be times when the available tasks will not comply with their individual preferences. \\
\textit{So that is what I have communicated with my team, that all right you get exciting things to do but there are times when you need to work on a repetitive chunk and that’s just part of the project stream, you can't avoid that.  So I think they understood that and they appreciate the nature of the project as well, and when an exciting challenge comes along they're all up for it, all right, okay, I want to do that as well, I want to work on these things. \textbf{P35~(manager)}}\\
Managers may need to disregard developers' preferences and goals in certain situations, but there were very few managers who did this regularly. Managers understand that if the developers are not excited about what they are doing, productivity and developer turnover will be impacted. 

\subsection{Managers Approaches}
\label{sec:approaches}
 
Examining the strategies presented in the previous section, we recognized that managers had different approaches based on their risk tolerance. No managers purposely risked quality, cost, or outcomes to provide learning opportunities. Rather, the managers considered risk in different ways. We classify the managers' approaches as:
\begin{enumerate}
    \item \textbf{risk-averse}: managers are reluctant to offer opportunities  to  learn  new  technology, skills, tools, and domains,  preferring developers to stick to their area of expertise;
    \item \textbf{risk-balancing}: managers are reluctant to provide developers opportunities  to  learn  new  technology,  tools, and domains, but they offer opportunities when there is low risk; and
    \item \textbf{growth-seeking}: managers seek growth opportunities and encourage team members to try out new things. They have high trust and confidence in the developers. 
\end{enumerate}

\smallskip

These three approaches often vary based on the situation and context, so a manager does not always use a single approach, but varies their approach based on the situation. For example, a manager who has two teams may use different approaches with each team due to differences in context, such as team composition or product maturity. Here we describe the three approaches in more detail using examples from the study.

\textbf{Risk-averse} managers prefer their team members to pick tasks that lie within their expertise and can be delivered on time, mitigating any delivery delays and failure risks. The manager wants the developers to understand the consequences of their self-assignment decisions. This could be because sometimes it is the manager, not the developer, who is held accountable for delays or quality issues.\\
\textit{I [lead] want to make sure that you [team member] are choosing work because you have confidence that you can apply your skills to getting it done. Otherwise you're just setting yourself up for a disaster and failure, and you're imposing risk on the broader context. So I want to make sure that that [task] is done. \textbf{P18~(manager)}}

Although theoretically, agile teams should be self-driven, self-organizing, and self-assigning, our results show this does not always occur. Managers often step in to facilitate or even enforce specific assignments. 
As an example, one manager stated a preference for people to choose tasks that they are skillful in and have some prior knowledge. In this case, the opportunity to learn new tools or technologies is denied in favor of quick task completion.\\
\textit{I [manager] would like them [team members] to consider their current knowledge level.  So, to one person, Ah that’s obvious. To another, what is this? So, I would like it if they could assign that way, so that they can get a good Sprint. Otherwise they end up spending the first week getting familiar with the service or the technology they're working with, only to find out that they now just have an inkling of what questions to start asking, let alone doing the work. \textbf{P19~(manager)}}
\par
The managers recognize that when a developer does not have the required expertise or experience, the task will take more time to accomplish. The assignee will have to get familiar with the technology, tools, or domain before coming up with a solution. Even if the task was completed on time, there is also a risk to the quality. A risk-averse approach does not provide opportunities to learn.

\textbf{Risk-balancing} managers recognize the importance of providing growth opportunities and considering developers' interests and preferences. However, they believe these must be balanced with the risks to cost or product quality. If there is some risk, then developers should stick to tasks where they have expertise. An example is :\\
\textit{ I would not care much how much somebody wishes to work with something. So my decision would be very objective. ....... I mean I do want people to grow, I would want people to work on new technologies, but I cannot let them do that on client’s expense or on compromising the product quality.... \textbf{P33~(manager)}} 

Risk-balancing managers can empower their team to practice self-assignment while considering the risks involved in each assignment. For example, a manager said: \\
\textit{ I [lead] don't want people to pick a piece of work and go ‘I don't have enough time to do this’. So, you know, you have a three-pointer [task] and it’s the last day of the Sprint. You’re just, you’re going to fail, right?  Why?  Because a three-pointer in our research showed that it was going to take you about two days to get it done so you have made bad choices through the Sprint.\textbf{ P18~(manager)}} 

We also saw cases where team members with little or no experience were forbidden by their risk-balancing manager to select multiple tasks. These managers want to make sure the team members choose an appropriate number and size of tasks that can be finished within the allocated time. To do this, they set strict limitations.\\ 
\textit{ I [manager] would also like that if they [team member] don't have experience, then they don't pick too much stuff for the Sprint, to make sure they don't underestimate how much time it's going to take. Coz I [manager] get very frustrated when somebody comes to the end of the Sprint, they've got three stories to do, none of them are done yet but `Yep, I'm [team member] all good'.  They're not good. \textbf{P19~(manager)}}\par 
There are also situations when the developers consider their own preferences without considering the value added to the project. An example is a very motivated novice developer who wants to use state-of-the-art tools and frameworks, ignoring the value that they may (or may not) add to the product (e.g., a throwaway prototype for a quick roadshow). In such situations, risk-balancing managers want the developers to realise the value they are expected to add for the customer. 

\textbf{Growth-seeking} managers encourage team members to step out of their comfort zone and learn new skills. For example:\\
\textit{When we're [team] sitting down committing to a Sprint, I'll [lead] just ask 'Okay so who's going to do something that's out of their comfort zone?'  And then somebody will go, 'hey, you know, this front-end task, I'm a back-end guy, and this front-end task looks like I'd be able to do it Or, I've been meaning to learn some new thing about CSS, here's an opportunity for me to do that'.  \textbf{P18~(manager)}} 
\par
Such managers are interested in the developer's growth and help them identify learning opportunities.\\ 
\textit{If I feel that there are bits of work that are better to help them, so I have one to ones with them. I'd like to understand where they would like to go. I sort of push them to say 'hey perhaps you might be interested in doing this one, this is gonna be a quite interesting piece of work if you wanna learn x, y, and z'. \textbf{P40~(manager)}}\\
Managers employing this approach are more likely to let the team members assess the risk because of their trust and confidence in them. 

\medskip

\noindent\fbox{\begin{minipage}{25.5em}
\textbf{Summary for RQ2:}  Both  managers  and  developers  are happy when developers self-assign tasks where they have prior experience or technical expertise and understand the task well. Conflicts  arise when developers  self-assign unfamiliar  tasks  motivated by a desire for  learning while ignoring  business  priorities. Managers balance  individual preferences and project goals using  different  strategies while applying risk-averse, risk-balancing, or growth-seeking approaches.
\end{minipage}}

\section{Discussion}
\label{sec:discussion}
Here, we present situations that managers should be aware of and present recommendations for their guidance. We then discuss how our findings relate to previous work, threats to validity, and future work.
\subsection{Implications: Recommendations for Managers }
In an agile environment, user stories and tasks are meant to be ordered by their priority. Still, both developers and managers acknowledge that developers tend to put their individual preferences over business priorities and needs. From our analysis, we identified specific situations that managers should be aware of, which we call \textit{red flags}. We also suggest guidelines for managers to deal with these situations.
\begin{itemize}
\item \textbf{\textit{Most developers are selecting only tasks related to their skill set:}} If developers keep working on tasks related to their skill set, then, in the long run, this will result in highly specialized team members, threatening cross-functionality. \\
\textit{Managers should intervene if they see some needed skills are limited to only certain developers. They should encourage developers to self-assign tasks outside their comfort zones to continue their growth and encourage cross-functionality.}
\item \textbf{\textit{Everyone selects low priority tasks of interest:}} If most of the developers are disregarding business priorities and important work isn’t being done, then customer needs will not be met. \\
\textit{Managers should monitor the priority of tasks being self-assigned. If most developers are ignoring the high priority tasks, the manager should remind them of the importance of delivering high business priorities.} 
\item \textit{\textbf{Senior developers take all the interesting tasks:}} Senior developers may select all of the exciting and challenging tasks since their expertise will bring time savings. However, if the junior developers are frequently working on boring and mundane tasks, it will reduce their opportunities to grow new skills and demotivate them eventually.\\
\textit{Managers should pay special attention to ensure junior developers get an equal share of interesting work and growth opportunities. Senior developers can work with junior developers to enable learning and knowledge sharing. Managers should incentivise both junior and senior developers. Instead of taking the interesting tasks for themselves, seniors should mentor juniors, and juniors should volunteer for interesting tasks.}
\item \textbf{\textit{Developers avoid time-pressing tasks:}} Many developers prefer not to select urgent or critical tasks with high severity (e.g., a critical update to a live product) as these tasks have high impact and can be stressful. If a small set of developers are always taking these tasks, it can lead to burnout.\\
\textit{Managers should ensure that the developers are not regularly or unduly stressed due to work load, complexity, competition, or pressure of performance. Everyone in the team should help with the critical and stressful tasks. Managers should share the significance and impact of tasks with the team so that developers can keep these into account while choosing tasks.}
\item \textbf{\textit{Developers avoid undesirable tasks:}} Developers avoid certain tasks intentionally due to personal dislikes. Verifying bugs in legacy systems with poor documentation or updating documentation are  generally undesirable to developers. This can be particularly problematic if everyone on the team finds a particular task undesirable as it can be left undone.\\
\textit{Managers need to keep an eye on the task board. They should give gentle reminders to the team members when some tasks remain unassigned for a long time, nudging them towards completion.} \textit{They can hold periodic triage meetings to go over long-lived tasks and push developers to finish them before a set time, e.g., some future release.}
\item  \textit{\textbf{Ignoring developers' personalities:}} Developers with certain personality traits (e.g., introverts) are more likely to defer to others or less likely to voice their individual preferences. If these behaviors are left unnoticed, it could be unhealthy for them in the long run.\\
\textit{Managers should be aware of developers' different personalities, aptitudes, and skill sets to help ensure fairness in work allocation. Managers could have one-on-one discussions with the developers to understand their personalities and aptitudes. They could encourage developers to take personality and skills assessments to help them be more aware of their own preferences, strengths, and weaknesses. Managers could mentor them to address their weaknesses. They could help them define their personal goals and help them choose tasks that align with their growth plan.}
\end{itemize}

We also identified some general recommendations for managers, based on the strategies reported in Section \ref{sec:strategies}, which can help them to facilitate and drive a sustainable self-assignment within their teams.  
\begin{itemize}
\item Managers should encourage developers to define their individual development goals and share them with the team and managers. 
\item Managers can introduce skill development and improvement programs within teams. For example, they can have a dedicated skill development day or knowledge sharing sessions.
\item Managers should keep an eye on the self-assignment choices the developers make. They should help developers understand and realise the consequences of their choices and help them find a balance across all factors.
\item Managers can introduce mentoring programs within their teams. This will motivate the developers to focus on their personal growth with some light-weight accountability.
\item Managers can introduce skill appreciation and endorsement programs to encourage and boost developer motivation. For example, managers can use social media platforms and team meetings to acknowledge and endorse developers' achievements. However, endorsements should be given in a way that is not detrimental to other developers' self-efficacy and well-being. Some developers might get more chances to improve their learning curve and show progression while others might not. Managers should treat each case individually keeping into account the nature of developers’ work to create a balance of appreciation.
\item Managers should create a friendly environment where the team feels comfortable discussing their issues and are not hesitant to ask for help. They should know they have support if they get stuck. In addition to creating a friendly environment, a manager should look for factors that indicate a healthy and safe environment. The managers should occasionally check the health of the team environment. A manager should identify and fix an environment that is hostile to feedback and discussion. To build synergy between developers and managers, managers should allow informal conversation over a cup of coffee or lunch and apply an open door policy to encourage open communication and feedback.
\end{itemize}

\subsection{Comparison with Related Work}

No other work specifically investigates factors developers consider while self-assigning tasks. We investigated such factors and found that many of them were related to human and social aspects. For example, factors such as the task's learning potential, the task's potential for recognition, and the task's desirability are related to developers' motivations.

Many software engineering research studies have focused on human and social aspects such as motivation~\cite{beecham2008motivation}, job satisfaction and perceived productivity~\cite{storey2019towards}, happiness~\cite{graziotin2017unhappiness}, emotions~\cite{murgia2014developers},  team-work and collaboration~\cite{herbsleb2006collaboration}, communication and coordination~\cite{barkhi2006study}, group awareness~\cite{gutwin2004group}, trust and knowledge sharing~\cite{pinjani2013trust}. Some of the factors we identified in this study are related to individual motivations, thus research on motivation is closely related to this study. In this section, we cover research into motivation in software engineering in general and then motivation in task allocation, and finally we compare our findings to other studies. 

Motivation is defined as "\textit{factors and events that drives human behaviour over time}". Researchers have studied the concept of motivation from different perspectives such as through different theories (e.g.,~\cite{ryan2000self,katzell1990work}) and across different fields (e.g.,~\cite{steers2004future}). Motivation has been related to various constructs such as job satisfaction, performance, enthusiasm, and quality of work. Many studies have found that motivation has a positive impact on software teams. Having motivated software engineers results in higher quality software, more successful projects, better productivity, improved task performance, and a greater sense of accomplishment~\cite{beecham2008motivation, hall2008we}. High motivation also results in decreased employee turnover, budget overflow, and delays in project delivery~\cite{francca2011motivation}.

Many studies have investigated what drives the motivation and satisfaction of software engineers and developers at work~\cite{hall2008we, beecham2008motivation, francca2011motivation}. Hall et. al found that autonomy, technically challenging work, work-life balance, variety of work, and rewards and incentives are motivators for software developers~\cite{hall2008we}. Literature on motivation and job satisfaction reports that the most frequently cited motivators for software engineers are identifying with the task, having a clear career path, and having a variety of tasks~\cite{beecham2008motivation, francca2011motivation}. They also found that high employee participation and good management are important motivators. A recent multi-case study investigated how work motivation and job satisfaction of software engineers are impacted by workplace factors~\cite{francca2018motivation}. The study found that well-defined work, cognitive workload, useful knowledge, work variety, creative and challenging work, accomplishment, and recognition lead to higher motivation. Developers are also motivated by solving problems, producing high quality work, refactoring, creating something new, and helping others~\cite{baltes2018towards}.

Studies have also explored factors that influence motivation in agile teams. One study conducted on agile teams found that factors, such as autonomy, variety, significance, feedback, and ability to complete a whole task, result in motivated and satisfied software developers in large agile projects, leading to lower turnover and higher job satisfaction~\cite{tessem2007job}. Another study that examined motivation through a case study of three agile companies found that developers in an agile context are motivated by opportunities to widen skills or try something new, a lack of bureaucracy in the development process, a feeling of accomplishment, and an elimination of waste~\cite{melo2012developers}. 
In summary, there is much evidence that motivation of software developers is important, and there are a variety of general motivators for software developers. 

Having motivated team members is a key principle of agile methods. The agile manifesto states \textit{build projects around motivated individuals}~\cite{de2018agile}.  Prior work has found that software developers are motivated by opportunities to learn, complex problems, and a variety of work~\cite{moe2008understanding}. If we look at agile methods such as Scrum, we know that developers are meant to select tasks based on priority and business value \cite{deemer2010scrum,sutherland2013scrum}. However, at the same time, Scrum encourages people to select tasks that promote learning. In practice, developers consider different factors when they self-assign tasks \cite{masood2020agile, masood2017motivation}. It gets challenging when team members regularly pick easier tasks or tasks outside their expertise \cite{hoda2016multi}. Team members need to balance between the freedom of choice with responsibility to sustain autonomy \cite{hoda2010balancing, moe2008understanding}. However, this is easier said than done and managers often need to facilitate this process. This study explores the manager's role in balancing individual preferences with business outcomes while self-assigning tasks. 

The role of the manager is well-explored in multiple studies in software development~\cite{shastri2017understanding,gandomani2020role,de2018agile}. These studies reported the different ways managers facilitate the teams as part of their day to day responsibilities. Kalliamvakou et.~al conducted a mixed-method empirical study to explore manager attributes~\cite{kalliamvakou2017makes}. A study at Google also reported a set of manager behaviors~\cite{garvin2013google}. These studies have explored the multi-faceted role of the manager. Some examples of manager attributes and behaviours from these studies are: \textit{good coach, empowers team, motivates the engineers, does not micro manage, good communicator, helps with career development, has a clear vision, possess technical skills that can help the team, and mediates communication}. On the other hand, our study focused on the agile practice of self-assignment and the role of the manager in facilitating a sustainable self-assignment. Our findings relate to some of the manager attributes and behaviors these studies uncovered. Managers encouraging the developers to self-assign tasks enables autonomy, empowers the team, and motivates the engineers.  Similarly, providing developers the opportunities to learn new technologies, skills, tools, or domains helps them develop their talent~\cite{kalliamvakou2017makes, garvin2013google}. Additionally, our study identified different approaches managers adopt and their strategies to balance individual preferences and business outcomes.

Providing learning opportunities is acknowledged as a need for happy, motivated developers~\cite{melo2012developers}. Studies have reported practical actions that managers can take to increase developer motivation by providing them promising learning avenues. Some previous recommendations like praising developers' efforts (`\textit{supervisory encouragement}') and giving developers autonomy (\textit{`freedom'}) are in line with our findings~\cite{van2000process}. Developers change their teams to improve opportunities for learning new things~\cite{hilton2018study}. However, these are generally advised for software developers irrespective of what software development methodology they apply. We have developed a set of detailed recommendations specific to agile methods and self-organizing teams. 

\subsection{Threats to Validity \& Future Work}
We describe potential threats to validity and how they were mitigated by considering \textit{reliability}, \textit{construct validity},\textit{ internal validity}, and \textit{external validity} threats following the guidelines proposed by Runeson et al.~\cite{runeson2009guidelines}. 

The \textit{reliability} of our findings can be impacted by researcher bias, researcher error, and participant bias. Thus, we took care to consider these threats throughout the study. To mitigate researcher bias, extensive discussions were held between all authors on the data collection, analysis, and results to ensure mutual consensus, understanding, and cross-verification. The first author generated the codes, concepts, and categories and identified the relationships evidenced in the data while applying open and axial coding. The second and third authors reviewed the results and validated the coding procedures during discussion meetings throughout the research study. The researcher conducted a maximum of two interviews per day to mitigate researcher fatigue and any resulting researcher error threat. To minimize participant bias, we adopted several mitigation strategies. First, we ensured that participants understood that their identity would not be revealed in any resulting publications to encourage honest conversations. We shared the research study details including objectives, potential risks, participant information sheets, and consent forms with the participants before data collection. We also scheduled the interviews based on the participants' preferences to ensure their comfort. 

To address the threat of \textit{construct validity}, the instruments used for data collection (interview guides and pre-interview questionnaires) were developed, reviewed, and revised iteratively by the research team throughout the study. Also, a pilot interview was conducted to gauge the interview's duration, understandibility of the questions, and coverage of the research scope. To mitigate the threat of \textit{internal validity} and misrepresentation, we included only study participants who use agile methods and practice self-assignment by collecting details of agile experience and use of practices in a pre-interview questionnaire. 
Regarding threats relating to \textit{external validity} and \textit{generalizability} of the research findings, we recruited participants through multiple channels such as social media, networking platforms, and personal references. Our participants were diverse in regards to ethnicity, country, gender, age, experience, role, company size, and project domain. Still, we do not claim generalizability due to the nature of the study. Our data set is not representative of the entire international agile community. However, we did have a suitable number of participants for a qualitative study \cite{creswell2007qualitative,denzin2005discipline}, and we achieved theoretical saturation.

Our findings pave the way for \textit{future work} in this area. For instance, we noticed that some of the factors appear to be related to motivation, e.g., learning potential and task desirability. On the other hand, factors such as the opinion of the manager or business priority are more like constraints or requisites enforced by the system, rules, values, or principles.  Also, we noticed that motivations behind reported factors like completion time and previous experience may or may not be based on personal preferences or team goals. These can vary with different situations and contexts as indicated through examples in Section \ref{sec:tradeoffscontext}. Future studies can further explore the factors we have identified and the relationships between the different factors, investigating the link between motivation and self-assignment. Future work can investigate interesting situations such as internal politics~\cite{magazinius2012investigating} that could influence self-assignment. A quantitative survey on the relative importance of the reported factors could validate our findings as well as enable more detailed analysis on the relationship between the factors and different demographics like company sizes, industry domains, gender, experience, or age. This could lead to more fine-grained and personalized recommendations for managers and developers. Further, the impact of the factors on job satisfaction, productivity, career growth, and software quality can be studied.    
It is not uncommon that managers stereotype developers expertise by their past experience. It is challenging for the developers to work on new goals to acquire new skills. This is an ongoing challenge software researchers and designers need to address while designing and developing automated recommendation-based tools. Finally, our findings can be validated in other settings such as through studies of the Open Source Software~(OSS) communities.

\section{Conclusion}
\label{sec:conclusion}
We found three categories of factors for task self-assignment: \textit{task-based, developer-based}, and \textit{opinion-based}. Some of these factors, like a \textit{task's learning potential} are prioritized by developers. While managers are aware of these preferences, they would prefer for developers to consider a task's \textit{business priority} and select tasks that align with a developer's \textit{technical ability} and \textit{previous experience}. Conflict arises when developers self-assign tasks based on its learning potential only, ignoring factors such as business priority, technical expertise, and understandability. We identified several strategies that managers employ to facilitate self-assignment and reconcile individual preferences with business priorities. We classified managers' approaches as \textit{risk-averse, risk-balancing}, and \textit{growth-seeking} approaches. From our findings, we created a set of guidelines that can help agile managers empower their teams to practice sustainable self-assignment. 

\ifCLASSOPTIONcompsoc
  \section*{Acknowledgment}
\else
  \section*{Acknowledgment}
\fi
We thank all participants who contributed to this study. This study was conducted under approval from the Human Participants Ethics Committee at the University of Auckland. We would also like to thank
our anonymous reviewers for their insightful comments and feedback that helped to improve the quality of our paper and research.

\bibliographystyle{plain}
\bibliography{SE_Factors.bib}

\begin{IEEEbiography}[{\includegraphics[width=1in,height=1.25in, clip,keepaspectratio]{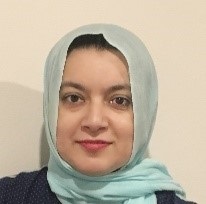}}]{Zainab Masood}is a final year PhD student at the University of Auckland, New Zealand. Her research interests include software development methodologies and practices, software testing and quality assurance, and human aspects of software engineering. She has published her research in the IEEE Transactions on Software Engineering, Empirical Software Engineering, and the Journal of Systems and Software. Contact her at zmas690@aucklanduni.ac.nz
\end{IEEEbiography}


\begin{IEEEbiography}[{\includegraphics[width=1in,height=1.25in,clip,keepaspectratio]{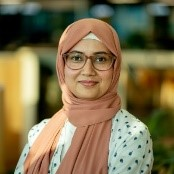}}]{Rashina Hoda}is an Associate Professor in Software Engineering at the Faculty of Information Technology, Monash University, Melbourne. Her research focuses on human and social aspects of software engineering, socio-technical grounded theory, and serious game design. She serves on the IEEE Transactions on Software Engineering review board, IEEE Software advisory board,
as ICSE2021 social media co–chair, CHASE
2021 program co–chair, and ICSE2023 SEIS co-chair. For details see www.rashina.com. Contact her at
rashina.hoda@monash.edu
\end{IEEEbiography}


\begin{IEEEbiography}[{\includegraphics[width=1in,clip,keepaspectratio]{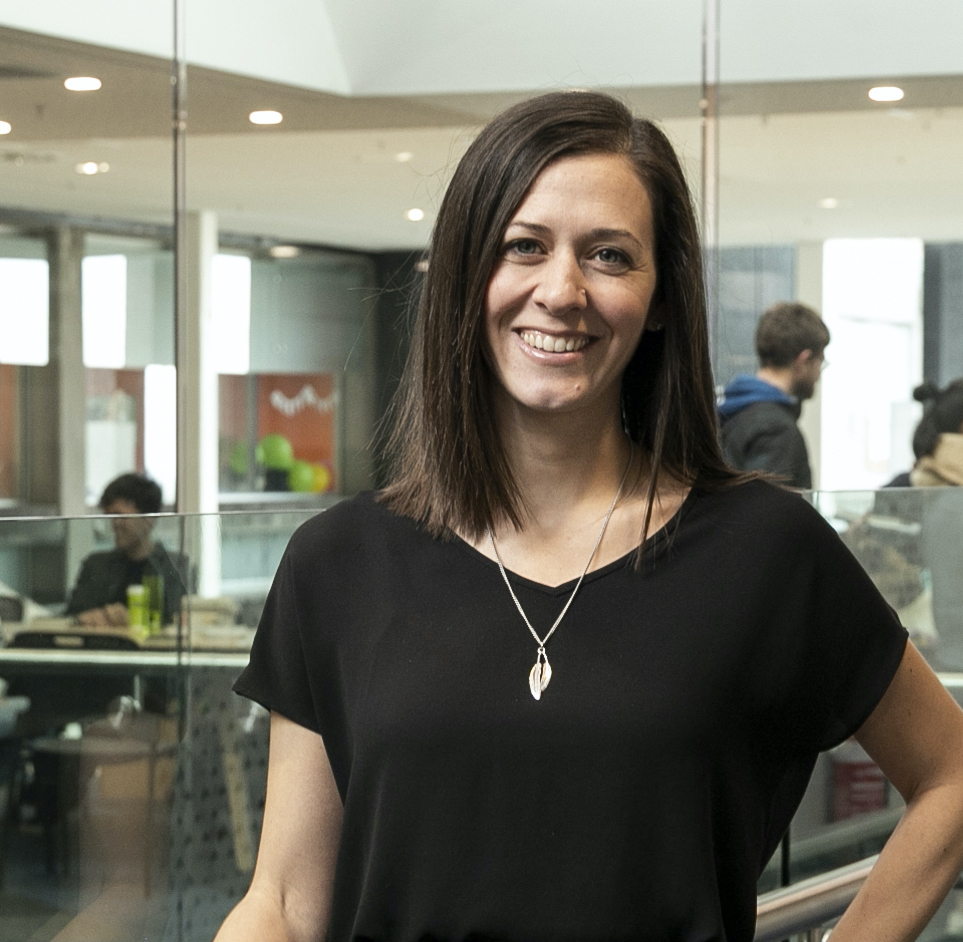}}]{Kelly Blincoe}is a Senior Lecturer in Software Engineering at the University of Auckland, where she leads the Human Aspects of Software Engineering Lab (\url{https://hasel.auckland.ac.nz/}). Her research areas include collaborative software development, software analysis, and software requirements. She currently serves on the editorial board of the IEEE Transactions on Software Engineering, the Empirical Software Engineering Journal, and the Journal of Systems and Software. She is also on the Executive Board of Software Innovation New Zealand. Contact her at k.blincoe@auckland.ac.nz
\end{IEEEbiography}

\end{document}